\documentclass[aps,prb,twocolumn,amsmath,amssymb]{revtex4}
\usepackage{graphicx}
\usepackage{dcolumn}
\usepackage{bm}
\usepackage{color}
\usepackage{here}
\definecolor{DarkBlue}{rgb}{0.1,0.1,0.5}
\definecolor{Red}{rgb}{0.9,0.0,0.1}
\definecolor{Green}{rgb}{0.0,0.99,0.0}

\begin{document}

\title{Momentum dependent light scattering in insulating cuprates}

\author{F. H. Vernay$^{1,2}$, M. J. P. Gingras$^{1,3}$, and T. P. Devereaux$^{1,2}$}

\affiliation{ $^{1}$Department of Physics and Astronomy,
University of Waterloo, Waterloo, Ontario N2L 3G1,
Canada\\
$^{2}$Pacific Institute for Theoretical Physics, University of British Columbia,
Vancouver, B. C. V6T 1Z1, Canada\\
$^{3}$Department of Physics and Astronomy, University of Canterbury, Christchurch 8020, New-Zealand}
\date{\today}
\begin{abstract} %
We investigate the problem of inelastic x-ray scattering in the
spin$-\frac{1}{2}$ Heisenberg model on the square lattice. We
first derive a momentum dependent scattering operator for the
$A_{1g}$ and $B_{1g}$ polarization geometries. On the basis of a
spin-wave analysis, including magnon-magnon interactions and
exact-diagonalizations, we determine the qualitative shape of the
spectra. We argue that our results may be relevant to help
interpret inelastic x-ray scattering experiments in the
antiferromagnetic parent cuprates.
\end{abstract} %
\maketitle The advances made in 3$^{rd}$ generation light sources
have recently provided new insights into the study of electron
dynamics in strongly correlated systems via resonant inelastic
x-ray scattering (RIXS). Detailed information has already been
obtained on Mott gap excitations and orbital transitions as a
function of doping in the cuprate
families\cite{Mott,orbit,mott2,mott3,mott4,kotani}. However, due
to limitations in resolution, $\sim 300$ meV near the elastic
line, excitations at low energies remain hidden. On the other
hand, low energy excitations at small photon momentum transfers,
e.g., phonons, magnons, and electron-hole excitations in metals
and superconductors, have been well characterized via Raman
spectroscopy\cite{RMP}. Yet, achieving a detailed understanding of
the momentum  and polarization dependence of low energy
excitations in strongly correlated matter would greatly clarify
the interplay between various degrees of freedom, such as
antiferromagnetism, charge density wave order, and
superconductivity, often present in system with many competing
interactions\cite{Dagotto}.

In the energy range below a few hundred meV lies one of the most
prominent features observed via Raman scattering in Heisenberg
antiferromagnets - the two-magnon feature at energy $\sim 2.7J$,
with $J$ the nearest-neighbor magnetic exchange. While the
peak frequency of the broad two-magnon peak in Raman scattering is well
understood, the asymmetry of the lineshape as well as the
polarization dependence remains unexplained, even though it has
been lavished with attention \cite{freitas}. Preliminary RIXS on
the Cu $K$-edge in LaCuO$_{4}$\cite{data1} and M-edge in CaCuO$_{2}$
\cite{data2} have shown evidence for low energy two magnon
scattering. Since in the near future this low energy window will
open for inelastic x-ray studies, it will afford an opportunity
to study the dynamics of magnon excitations via the charge
degrees of freedom which will complement neutron and Raman
scattering studies.

Lorenzana and Sawatzky\cite{lorenzana} investigated the
properties of phonon-assisted absorption in 2D Heisenberg
antiferromagnets, and investigated the bi-magnon spectrum probed
by infrared conductivity. In this paper we extend these
calculations to present a theory of inelastic x-ray scattering of
the two-magnon response in a Heisenberg antiferromagnet. In
particular we show that the momentum and polarization dependence
can provide detailed information on the nature of magnon-magnon
interactions in parent insulating cuprate compounds.

We remark at the outset that we neglect specific resonant matrix
elements relevant to the RIXS process including transitions
resulting from the creation of the core hole, and work in the
restricted model of the half-filled single-band Hubbard model on
the square lattice to capture scattering via magnon creation.
Thus we neglect specific pathways of charge excitation, such
as the Cu $K$-edge 1s-4p transition, and how double-spin flips
may occur near the site where the core hole is created.
Such kinematic details are indeed important to determine
accurately the RIXS intensity as well as the proper symmetry and
polarization of spin excitations which may be probed by specific
x-ray transitions. However, in order to obtain a preliminary
understanding of how the two-magnon response can be probed and
how polarization may enter, we first simply focus on the
evolution of the two-magnon response, far from any specific
resonance, for nonzero photon momentum transfers ${\bf q}$.

Transitions via light scattering can be created via dipole or
multipole matrix elements involving states within the conduction
band or out of the valence band. These transition may be selected
by orienting incident and scattering polarization light vectors,
${\bf {\hat e}}_{i}$ and ${\bf {\hat e}}_{f}$,  respectively. The
photons entering in the scattering process are represented by
$(\omega_{i,f},{\bf k}_{i,f},{\bf {\hat e}}_{i,f})$ where indices
$i$ and $f$ represent the incoming and outgoing photon,
respectively. Since we are interested in the insulating phase,
the scattering of light is induced by the interband part of the
operator. Following Refs.~[\onlinecite{loudon,shastry}],
we derive a finite momentum transfer ${\bf q}$, ${\bf q} \equiv
{\bf k}_i-{\bf k}_f$,
 scattering operator for different polarization
geometries. The interband part of the scattering matrix element
is given by:
\begin{equation}\label{scat}
\begin{array}{rcl}
\langle f|M_r|i\rangle &=&
 \sum_\nu \left[
\frac{\langle f|{\bf J_{k_f}}\cdot{\bf {\hat e}}_f|\nu\rangle
\langle\nu|{\bf J_{-k_i}}\cdot{\bf {\hat e}}_i|i\rangle}
{\epsilon_\nu-\epsilon_i-\omega_i}
\right.\\
&&+\left.
\frac{\langle f|{\bf J_{-k_i}}\cdot{\bf {\hat e}}_i|\nu\rangle
\langle\nu|{\bf J_{k_f}}\cdot{\bf {\hat e}}_f|i\rangle}
{\epsilon_\nu-\epsilon_i+\omega_f}\right]
\end{array}
\end{equation}
where $\nu$ represents states out of the lower Hubbard band, and
${\bf J_k}$ is the current operator ${\bf J_k}=\sum_{\bf p,
\sigma} \frac{\partial\epsilon_{\bf p}}{\partial {\bf p}} ~
c^\dagger_{{\bf p+\frac{k}{2}},\sigma}c_{{\bf
p-\frac{k}{2}},\sigma}$, with $\epsilon_{\bf p}=-2t\left[\cos(p_x
a)+\cos(p_y a)\right]$ for a square lattice with lattice constant
$a$, and nearest-neighbor hopping $t$. The current may be
expressed as:
\begin{equation}\label{current}
\begin{array}{rcl}
{\bf J_{k_f}}\cdot{\bf {\hat e}}_f
&=& it \sum_{\bf r,\bm\delta,\sigma}  {\hat e}_f^{\bm\delta}
e^{-i{\bf k_f}\cdot({\bf r+\bm\delta}/2)}\\
&\times& \left[c^\dagger_{{\bf r+\bm\delta},\sigma}c_{{\bf r},\sigma}-
c^\dagger_{{\bf r},\sigma}c_{{\bf r+\bm\delta},\sigma}\right]\\
&\equiv& \sum_{\bf r,\bm\delta,\sigma} {\hat e}_f^{\bm\delta} J_{\bf k_f}({\bf r,\bm\delta}),
\end{array}
\end{equation}
where ${\hat e}_f^{\bm\delta}$ is the projection of the
polarization vector along the neighbor direction ${\bm \delta}$.
In the insulating state, $|i\rangle$ and $|f\rangle$ are single
occupied states, whereas $|\nu\rangle$ are excited states
consisting of a doubly occupied state at site ${\bf r+{\bm
\delta}}$. Thus the energy of the excited state is roughly given
by $\epsilon_\nu\approx U+\epsilon_i$, and the intermediate
states may be collapsed using the identity $\frac{1}{4}-{\bf
S}_i\cdot{\bf S}_j=\frac{1}{2} c^\dagger_{i,\sigma}c_{j,\sigma}\
c^\dagger_{j,\sigma'}c_{i,\sigma'}$, which simply means that the
exchange process (and therefore a doubly occupied site) is
allowed only if  spins $\sigma$ and $\sigma'$ are
opposite\cite{shastry}. Finally, we recast $\langle f\vert
M_r\vert i\rangle$ in Eq. 1 via the following scattering operator:
\begin{equation}
\begin{array}{rcl}
{\mathcal O}({\bf q}) &=& 8t^2\sum_{\bf r,\bm{\delta}} e^{i{\bf
q}\cdot({\bf r+\bm{\delta}}/2)}
(\hat{\bm\delta}\cdot {\bf {\hat e}}_f)(\hat{\bm\delta}\cdot {\bf {\hat e}}_i)\\
&\times& \left(\frac{1}{4}-{\bf S_r}\cdot{\bf S_{r+\bm{\delta}}}\right)
\left[\frac{1}{U+\omega_f}+\frac{1}{U-\omega_i}\right].
\end{array}
\end{equation}
For crossed polarizations, ${{\bf \hat
e}}_{i,f}=\frac{1}{\sqrt{2}}({\hat x}\pm{\hat y})$, transforming
as $B_{1g}$, and for parallel polarizations along ${\hat x}$ and
${\hat y}$, transforming as $A_{1g}$, we obtain the following
expression~:
\begin{equation}\label{HLFB1g}
\begin{array}{lcr}
&&{\mathcal O}_{A_{1g},B_{1g}}({\bf q}) = -8t^{2}
\left[\frac{1}{U+\omega_f}+\frac{1}{U-\omega_i}\right]\\
&&\times\sum_{i,\bm{\delta}} P_{A1,B1}(\bm\delta) \, [{\bf
S}_i\cdot {\bf S}_{i+{\bm\delta}}]\, \cos ({\bf q\cdot
r_{i}}+{\bf q\cdot\frac{\bm{\delta}}{2}}),
\end{array}
\end{equation}
with $P_{A1,B1}(\bm\delta)$=1, for $\bm{\delta}=a\bm{\hat x}$,
and equals 1, -1 for $\bm{\delta}=a\bm{\hat y}$ for $A_{1g}$ and
$B_{1g}$, respectively. This is a finite momentum generalization
of the usual Loudon-Fleury light scattering operator, and for
${\bf q}=0$ the above expressions give the standard Raman
results~\cite{loudon,shastry}. We note that in general
subtractions of spectra for different polarization orientations
are needed in order to fully extract symmetry deconvolved
spectra\cite{RMP}.

Henceforth, we restrict ourselves to the half-filled Hubbard model
with $t/U$ small, and focus on its spin-$\frac{1}{2}$
antiferromagnetic Heisenberg representation with Hamiltonian
${\cal H}$. We first investigate the ${\bf q}$-dependent
inelastic response via the spin-wave (SW) approximation. We
proceed as usual
\cite{anders,katanin,kampf,CG,CF,davies,freitas,shastry} and
express ${\cal H}$ in its SW representation,
${\mathcal H}_{\rm SW} = Cst +\sum_{\bf k} \omega_{\bf k}
\left(\alpha_{\bf k}^\dagger\alpha_{\bf k}+\beta^\dagger_{\bf k}
\beta_{\bf k}\right)$, with $\omega_{\bf
k}=JSZ\sqrt{1-\gamma_{\bf k}^2}$ the magnon dispersion, and
$2u_{\bf k}^2-1=2v_{\bf k}^2+1= 1/{\sqrt{1-\gamma_{\bf k}^2}}$
and $\gamma_{\bf k}= (1/2)\left[\cos (\rm{k_x}a)  +\cos
(\rm{k_y}a)\right]$. Performing the same transformations to the
scattering operators ${\mathcal O}_{A_{1g},B_{1g}}$ we obtain:
\begin{equation}
\begin{array}{rcl}
&{\mathcal O}_{A_{1g},B_{1g}}&\propto \sum_{\bf k}
\beta^\dagger_{{\bf k}}\alpha^\dagger_{{\bf k}+{\bf q}}
\left\{-\left[\cos(\frac{{\rm q_x}a}{2})\pm\cos(\frac{{\rm q_y}a}{2})\right]\right.\\
&&\times(u_{\bf k}v_{{\bf k}+{\bf q}}+v_{\bf k}u_{{\bf k}+{\bf
q}}) +
(u_{\bf k}u_{{\bf k}+{\bf q}}+v_{\bf k}v_{{\bf k}+{\bf q}})\\
&\times&\left.\left[\cos({\rm k_x}a)\cos(\frac{{\rm
q_x}a}{2})\pm\cos({\rm k_y}a)\cos(\frac{{\rm q_y}a}{2})\right]
\right\},
\end{array}
\label{operator}
\end{equation}
where we have neglected the prefactor in Eq.~\ref{HLFB1g}.
For $A_{1g}$ and ${\bf q}=0$, we recover the familiar form of the
Raman operator which, being proportional to ${\cal H}$, commutes
with ${\cal H}$, giving no Raman scattering in that channel.

One also finds
${\mathcal O}_{A_{1g},B_{1g}}$ vanish
for
${\bf q}={\bf Q}=(\pi,\pi)$ and symmetry related points for both
$A_{1g}$ and $B_{1g}$. This is a consequence of including only
nearest neighbor spin interactions in the Heisenberg model
\cite{vdb}. While in our case the response vanishes for the
antiferromagnetic reciprocal lattice vectors, if one includes
longer range interactions, then this restriction may be
lifted\cite{CG}. Thus the x-ray Raman response for these
wavevectors may provide a window to sample the role of
longer-range spin interactions.

The scattering intensity
is proportional to $\langle f|{\mathcal O}_{B_{1g}}|i\rangle^2$,
while satisfying energy conservation $\omega=\omega_{\rm
k}+\omega_{{\rm k+q}}$. We first focus on the $B_{1g}$ channel,
where the two-magnon scattering is most prominent for ${\bf
q}=0$. The left panel of Fig. 1 shows the $B_{1g}$ bare scattering
intensity, neglecting magnon-magnon interactions: $I_{0}({\bf
q},\omega)=\frac{1}{2N}\sum_{\bf k}\delta(\omega_{{\bf
k}}+\omega_{{\bf k}+{\bf q}}-\omega)b_{{\bf k},{\bf q}}^2 $,
where $b_{\bf k,q}$ is the term in curly brackets in Eq. 5. The
$I_0({\bf q}=0,\omega)$ intensity recovers the standard Raman
response, which for the non-interacting case has a peak at
$\omega=4J$ due to the large magnon density of states at ${\bf
k}=(\pi,0)$ projected out by the $B_{1g}$ operator. For nonzero
${\bf q}$, two magnons are created with wavevectors ${\bf k}$ of
different magnitude and direction, leaving behind a reorganized
spin configuration, and the response is given by convolving two
magnon density of states at different ${\bf k}$ separated by
${\bf q}$. We note that while
the response is still governed by $\sim 4J$ for finite ${\bf q}$
due to the magnon density of states being largest at ${\bf
k}=(\pi,0)$, the intensity is highly suppressed
as ${\bf q}$ approaches ${\bf Q}$.

\begin{figure}[ht]
\begin{center}
\includegraphics*[scale=0.65,angle=0]{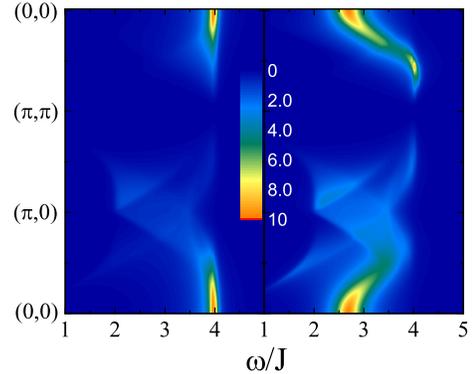}
\caption[figure1]{\label{SW} Spin-wave intensity spectrum without
($I_{0}({\bf q},\omega)$, left) and with ($I_{\rm RPA}({\bf
q},\omega)$, right) magnon-magnon interaction. The color
intensity scale is 2.2 times smaller for the right panel than the
left.}
\end{center}
\end{figure}

Magnon-magnon interactions need to be included to describe more
realistically the local spin rearrangement
\cite{kampf,katanin,anders,shastry,davies,CG,CF,freitas}.
In particular, interactions lead
 to a reduction of the peak frequency for $B_{1g}$
Raman to $2.78J$, where $J$, as shown by Singh {\it et al.}
\cite{singh}, is also renormalized by quantum corrections. Many
diagrams contribute to the magnon-magnon vertex corrections, and
a subset has been investigated for ${\bf q}=0$ Raman scattering
\cite{CG}. In that case, magnon-magnon interactions lower the
relevant energy scale from $4J$ to $3J$ due to the local breaking
of six exchange bonds for two neighboring spin flips. For finite
${\bf q}$ considered here, the two magnons are created with net
momentum ${\bf q}$ which distribute the spin flips over longer
length scales. Since magnon-magnon vertex corrections are
expected to weaken as the spin arrangement occurs at larger
length scales, we approximate the renormalized response by a
generalized random-phase approximation (RPA) form $I_{\rm
RPA}({\bf q},\omega) \sim \frac{I_{0}({\bf q},\omega)} {1+J({\bf
q})I_{0}({\bf q},\omega)}$, where the function $J({\bf
q})=[\cos({\rm q_x} a/2)+\cos({\rm q_y} a/2)]/2$ is taken to
recover the Raman form at ${\bf q}=0$ and properties of the
solution to the Bethe-Salpeter equation for finite ${\bf q}$. A
proper treatment of magnon-magnon interactions for all ${\bf q}$
in the SW framework is a topic of future research. As shown in
the right panel of Fig.~\ref{SW}, the magnon interactions for
larger ${\bf q}$ bring the peak down to $\omega=2.78J$ for
$B_{1g}$ ${\bf q}=0$ Raman, but the general weakening of the
magnon interactions at larger ${\bf q}$ gives back the
unrenormalized response at larger ${\bf q}$ with, in particular,
the response still vanishing at ${\bf q}={\bf Q}$. On the other
hand, the dispersion of the peak changes dramatically when
magnon-magnon interactions are included, where the peak hardens
at finite ${\bf q}$ from the $\Gamma$-point, ${\bf q}=0$.

\begin{figure}[ht]
\begin{center}
\includegraphics*[scale=0.2,angle=0]{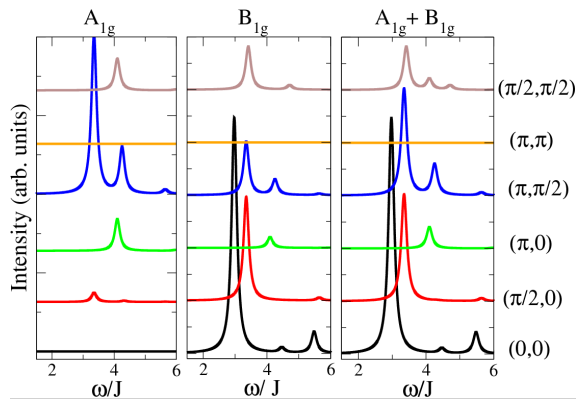}
\caption[figure2]{\label{ED_polar} 16-site cluster : exact
diagonalizations for $A_{1g}$ and $B_{1g}$ polarizations, to be
compared with spin-wave results. The spectra are shown for the
independent momenta ${\bf q}$ allowed for the $N=16$ site cluster
(given on the side of rightmost panel). The relative intensity of
the curves is given by $\langle {\mathcal O}({\bf q})^2\rangle$.}
\end{center}
\end{figure}

The results above differ slightly from
those obtained in Ref.~[\onlinecite{lorenzana}] for infrared
absorption from multi-magnons. While the dispersion of the peaks
is similar to our results, Ref.~[\onlinecite{lorenzana}] found a
bi-magnon response which is dominated by a sharp peak with a
small bi-magnon lifetime for momentum transfers $(\pi,0)$. Since
the form factors are different for infrared and polarized x-ray
Raman measurements, the bi-magnon spectrum has different
projections for the results presented here, favoring ${\bf q}=0$
for bi-magnon in $B_{1g}$ channel, and a growth of $A_{1g}$
component with increasing $q$. Thus the $A_{1g}$ response is more
similar to the infrared one, as it weights out similar regions in
the magnetic Brillouin zone (BZ). We note that phonon-assisted
bi-magnon response for x-ray Raman, considered in the same vein
as Ref.~[\onlinecite{lorenzana}], would also have similar
polarization dependent form factors as those considered here,
coming from the symmetry classification of the active phonon modes
involved in the scattering.

In order to explore the semi-quantitative validity of the SW/RPA
results, we investigate the ${\bf q}$-dependent inelastic x-ray
scattering for the Heisenberg spin Hamiltonian ${\cal H}$, using
an exact diagonalization approach, considering both the $A_{1g}$
and $B_{1g}$ channels. Although exact-diagonalization is limited
because of the prohibitive size of the Hilbert space, Gagliano
and Balseiro\cite{gagliano} demonstrated that it is a powerful
technique that allows to compute the dynamical quantities easily.
Noteworthy in the context of numerical investigations of Raman
spectra, Sandvik {\it et al.}\cite{anders} showed that even
though the spectra obtained by the Lanczos method are extremely
size-dependent, they are of direct relevance to test the response
calculated from a spin-wave analysis. As in
Ref.~[\onlinecite{dagotto}], the spectra are evaluated by
computing the continued fraction:

\begin{equation}\label{lanczos}
I({\bf q},\omega)=-\frac{1}{\pi}{\rm Im}\langle \Psi_{g.s.}|
{\mathcal O}^\dagger(-{\bf q})\frac{1}{z-{\mathcal H}} {\mathcal
O}({\bf q})|\Psi_{g.s.}\rangle
\end{equation}
with $z=\omega+i\epsilon+E_0$, where $E_0$ is the ground-state
energy, $\epsilon$ is a damping factor.

\begin{figure}[t]
\begin{center}
\includegraphics*[scale=0.28,angle=0]{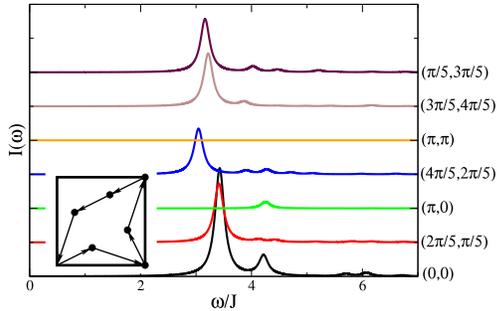}
\caption[figure3]{\label{ED_20} 20-site cluster : exact
diagonalizations in the $B_{1g}$ channel. The inset is a schematic
of the investigated ${\bf q}$-points.}
\end{center}
\end{figure}

We first compute the ground-state energy.
For $N=16$ sites we have $E_0/(NJ) \approx -0.70178020$ and wave-vector
$|\Psi_{g.s.}\rangle$.
We then begin evaluating the
continued fraction with the starting state:
\begin{equation}\label{state}
|\Phi({\bf q})\rangle=\frac{{\mathcal O}({\bf
q})|\Psi_{g.s.}\rangle} {\sqrt{\langle \Psi_{g.s.}| {\mathcal
O}^\dagger({\bf -q}){\mathcal O}({\bf q})|\Psi_{g.s.}\rangle}}
\end{equation}
Here, compared to Raman scattering, care has to be taken in
computing Eq.~(\ref{lanczos}): the state $|\Phi({\bf q})\rangle$
 being in a different subspace at nonzero ${\bf q}$ than the
ground-state, the matrix elements of ${\cal H}$ to compute the
continued fraction have to be expressed in the corresponding
subspace. The results are summarized in Figs.
\ref{ED_polar}-\ref{ED_20}.

For the case of $B_{1g}$ scattering, we recover prior results for
${\bf q}=0$\cite{shastry}, while we see that the peak disperses to
higher energies, approaching $4J$ both for ${\bf q}$
along the BZ diagonal as well as along the axes. In light of the
RPA results in the right panel of Fig. 1, we attribute this
dispersion as a weakening of magnon interactions at larger ${\bf
q}$.
At the same time, the overall intensity diminishes for larger
${\bf q}$
due to the form factor appearing in Eq. (\ref{HLFB1g}).

Since exact diagonalizations deal with small clusters, it is very
difficult to make a quantitative finite-size scaling analysis for
the spectral shape\cite{anders}. Therefore we repeat the
calculation for the $B_{1g}$ polarization with a 20-site cluster,
and plot the results in Fig.~ \ref{ED_20}. It is well known that
the 16-site cluster has additional symmetries
(hypercube-like)\cite{hasegawa} and it therefore important to
check for similar ${\bf q}$-points in the 20-site cluster. Due to
the different shape and boundary conditions of this cluster, we
are not able to explore the same momenta in the BZ. However
direct inspection of the ${\bf q}$ points $(0,0)$, $(\pi,0)$ and
$(\pi,\pi)$ reveal
that many of the features are qualitatively similar: the ${\bf
q}$=0 peak is at lower frequency than ($\pi,0$), and the
intensity weakens at finite ${\rm q}\equiv |{\bf q}|$ as in the
16-site case (compare with the middle panel of
Fig.~\ref{ED_polar}). In addition, the dispersion along the path
in the ${\bf q}$-space is of the same order $\sim J$. In the RPA
calculation, the minimum frequency of the main peak
is at $(0,0)$. For the compatible ${\bf q}$-points, the results
are consistent with the RPA approach as well (same qualitative
dispersion and minimum). We therefore believe that exact
diagonalizations give a correct qualitative description of the
response.

For $A_{1g}$, a peak appears for finite ${\bf q}$~\cite{vdb}.
Generally the peak disperses towards 4J for larger ${\bf q}$,
 following essentially the $B_{1g}$ spectrum. The $B_{1g}$
intensity
falls with increasing ${\bf q}$ while at the same time the
$A_{1g}$ spectral intensity grows. In fact, we note that the
intensity for $A_{1g}$ approaches that of the $B_{1g}$ spectrum
for ${\bf q}=(\pi,\pi/2)$, where the scattering operators in
Eq.~(\ref{HLFB1g}) are the same.


Thus, the overall intensity in an unpolarized measurement, given
by $I_{A_{1g}}+I_{B_{1g}}$ as shown in Fig. \ref{ED_polar}, is
dominated by the $B_{1g}$ channel for small ${\bf q}$, and a
mixture of $A_{1g}$ and $B_{1g}$ for larger $q$ ~\cite{B2g}. The
main results are thus (1) for small ${\bf q}$, the two-magnon
Raman peak rapidly disperses upward in energy, both for ${\bf q}$
along the BZ diagonal and BZ axes. At the same time, a
contribution from the $A_{1g}$ channel develops, as the intensity
of the $B_{1g}$ falls with increasing ${\bf q}$. We note, however,
that the problems of describing the two-magnon profile for ${\bf
q=0}$, which includes the differences between $A_{1g}$ and
$B_{1g}$ intensities, as well as the width of the lineshapes,
remains an issue here as well (see, e.g., Ref.
\onlinecite{freitas}). Thus more detailed treatments including
further spin-spin interactions, ring exchange, and possibly
spin-phonon coupling, would need to be incorporated in low energy
inelastic x-ray scattering as well. Besides affecting the overall
lineshape, the behavior at different momentum points, such as the
special point at $(\pi,\pi)$ may change. Thus, the detailed
momentum dependence of the spectra may be able to provide
important information of the types and extensions of spin
interactions in antiferromagnets.

Finally we remark on our results in terms of the current
capabilities of RIXS experiments. Empirically, from fitting the
two-magnon position and spin wave velocity, $J$ in the cuprates
is estimated to $J\approx 0.13$ eV.  Thus the two-magnon Raman
peak would lie generally obscured in the elastic line at currently
available resolution, which is several orders of magnitude larger
than the inelastic contribution. It is a possibility that the
two-magnon contribution would emerge from underneath the elastic
line at larger momentum transfers ${\bf q}$ as the peak in x-ray
Raman disperses to $4J$. If the energy resolution can be
enhanced, we propose that perhaps a clear two-magnon contribution
will become visible. However, a proper treatment of the resonant
matrix elements needs to be considered, which, while not
affecting the dispersion, may change the relative intensity of
the spectra at specific ${\bf q}$. This remains a topic for
future research. \cite{ack}

{\it Note added in proof:} After completion and submission of our
work we became aware of a closely related  preprint by Donkov and
Chubukov (cond-mat/0609002) The results are similar at small
momenta but they differ for  $\bf{q}=(\pi,\pi)$ due to a slight
difference in the form of their scattering operator.



\begin{thebibliography}{99}
\bibitem{Mott}
L. Lu, G. Chabot-Couture, X. Zhao, J. N. Hancock, N. Kaneko, O.
P. Vajk, G. Yu, S. Grenier, Y. J. Kim, D. Casa, T. Gog, and M.
Greven, Phys. Rev. Lett. {\bf 95}, 217003 (2005).

\bibitem{orbit}
P. Abbamonte, C. A. Burns, E. D. Isaacs, P. M. Platzman, L. L.
Miller, S. W. Cheong, and M. V. Klein, Phys. Rev. Lett. {\bf 83},
860 (1999).

\bibitem{mott2}
M. Z. Hasan, E. D. Isaacs, Z.-X. Shen, L. L. Miller, K. Tsutsui,
T. Tohyama, and S. Maekawa, Science {\bf 288}, 1811 (2000).

\bibitem{mott3}
Y. J. Kim, J. P. Hill, C. A. Burns, S. Wakimoto, R. J. Birgeneau,
D. Casa, T. Gog, and C. T. Venkataraman, Phys. Rev. Lett. {\bf
89}, 177003 (2002).

\bibitem{mott4}
Y. J. Kim, J. P. Hill, G. D. Gu, F. C. Chou, S. Wakimoto, R. J.
Birgeneau, Seiki Komiya, Yoichi Ando, N. Motoyama, K. M. Kojima,
S. Uchida, D. Casa, and T. Gog, Phys. Rev. B {\bf 70}, 205128
(2004).

\bibitem{kotani}
A. Kotani and S. Shin, Rev. Mod. Phys. {\bf 73}, 203 (2001)

\bibitem{RMP}
T. P. Devereaux and R. Hackl, Rev. Mod. Phys. {\it in press};
cond-mat/0607554.

\bibitem{Dagotto}
E. Dagotto, Science {\bf 309}, 257 (2005).

\bibitem{freitas} P. J. Freitas and R. R. P. Singh, Phys. Rev. B {\bf 62},
5525 (2000), and references therein.

\bibitem{data1}
J. P. Hill, Y.-J. Kim, private communication.

\bibitem{data2}
B. Freelon {\it et al.}, preprint.

\bibitem{lorenzana}J. Lorenzana and G. A. Sawatzky, Phys. Rev. Lett.
{\bf 74}, 1867 (1995); Phys. Rev. B {\bf 52}, 9576 (1995).
\label{Llorenzana}

\bibitem{loudon} P. A. Fleury and R. Loudon, Phys. Rev. {\bf 166}, 514 (1968)
\label{Lloudon}

\bibitem{shastry} B. S. Shastry and B. I. Shraiman, Int. Jour. Mod. Phys. {\bf B5}, 365 (1991)~;
B. S. Shastry and B. I. Shraiman, Phys. Rev. Lett. {\bf 65}, 1068 (1990)
\label{Lshastry}

\bibitem{anders} A. W. Sandvik, S. Capponi, D. Poilblanc and E. Dagotto,
Phys. Rev. B {\bf 57}, 8478 (1998)

\bibitem{katanin} V. Yu. Irkhin, A. A. Katanin, and M. I. Katsnelson,
Phys. Rev. B {\bf 60}, 1082 (1999)

\bibitem{kampf} A. A. Katanin and A. P. Kampf, Phys Rev B {\bf 67}, 100404(R) (2003)

\bibitem{davies} R. W. Davies, S. R. Chinn and H. J. Zeiger, Phys Rev B {\bf 4}, 992 (1971)

\bibitem{CG}
C. M. Canali and S. M. Girvin, Phys. Rev. B {\bf 45}, 7127 (1992).

\bibitem{CF}
A. V. Chubukov and D. M. Frenkel, Phys. Rev. B {\bf 52}, 9760
(1995).

\bibitem{vdb}
J. van den Brink, private communication.

\bibitem{singh}R. R. P. Singh, P. A. Fleury, K. B. Lyons and
P. E. Sulewski, Phys. Rev. Lett. {\bf 62}, 2736 (1989)

\bibitem{gagliano} E. R. Gagliano and C. A. Balseiro, Phys. Rev. Lett. {\bf 59},
2999 (1987)

\bibitem{dagotto} Elbio Dagotto, Rev. Mod. Phys. {\bf 66}, 763 (1994)
\label{Ldagotto}

\bibitem{hasegawa}Y. Hasegawa and D. Poilblanc,
Phys. Rev. B {\bf 40}, 9035 (1989)

\bibitem{B2g}
Here we neglect contributions from the $B_{2g}$ channel, which are
identically zero for nearest neighbor spin interaction.

\bibitem{ack} We acknowledge important discussions with B. Freelon, M.
Greven, M. Z. Hasan, J. P. Hill, Y.-J. Kim, M. V. Klein, S.
Maekawa, G. Sawatzky, K. M. Shen, Z.-X. Shen, and R. R. P. Singh.
Partial support for this work was provided by NSERC of Canada,
the Canada Research Chair Program (Tier I) (M.G), the Province of
Ontario (M.G.), Alexander von Humboldt Foundation (T.P.D.), and
ONR Grant N00014-05-1-0127 (T.P.D.). M.G. acknowledges support
from the Canadian Institute for Advanced Research.

\end{thebibliography}
\end{document}